# Scaling and universality in ontogenetic growth

Recently, West et al.[1] claimed to derive a general quantitative model based on fundamental principles for the allocation of metabolic energy between maintenance of existing tissue and the production of new biomass, and in addition claimed to derive a single, parameterless universal curve that describes the growth of many species. They further claimed that their model of the 3/4 exponent for the allometric scaling of metabolism provides a novel version of the growth equations and use the goodness of fit of growth-curve data to these equations as support of this interpretation. Here, we show that the universal curve arises from general considerations that are <u>independent</u> of the specific allometric model used and that the data do not distinguish between the 3/4 or the 2/3 exponent for the metabolic rate – mass scaling.

von Bertalanffy[2] (see his Equation (5)) considered a general equation of the form $dm/dt = a_\alpha m^\alpha - b_\alpha m^\beta$. The equation considered by West et al.[1] is a special case with $\alpha = 3/4$ and $\beta = 1$ while von Bertalanffy studied the case with $\alpha = 2/3$ and $\beta = 1$ in detail. Such an equation can be cast in a scaling form:

$$dm/dt = m^\alpha f(m/M), \qquad (1)$$

where $M$ provides a scale for the organism mass and is usually chosen to be the asymptotic maximum body size and $\alpha$ is an unspecified exponent. The scaling function $f$ approaches a constant value for small argument — when $m(t)$ is small, there is relatively little energy needed for the sustenance of the organism and essentially all the metabolic energy is funneled into growth processes. The requirement that $dm/dt = 0$ when $m$ reaches the value $M$ leads to the condition that $f(1) = 0$. Furthermore, one must ensure that the initial condition of the mass at $t = 0$ being equal to the birth-mass, $m_0$, is satisfied. For the von Bertalanffy[2] equation

$$f(x) = a_\alpha \left(1 - x^{1-\alpha}\right) \quad (2)$$

with $a_\alpha = b_\alpha M^{1-\alpha}$ to ensure that $f(1) = 0$.

We have analyzed the data from Ref. 1 on the cow, hen and guppy to assess whether one can discriminate between the two choices for $\alpha$. We find that Eqn. 5 of ref. (1) (which is a special case of Eq. 6 of ref. 2) written in the form

$$(m/M)^{1-\alpha} = 1 - \left[1 - (m_0/M)^{1-\alpha}\right] e^{-\gamma \alpha t} \quad (3)$$

or equivalently

$$r = 1 - e^{-\tau} \quad (4)$$

(with the dimensionless mass ratio $r = (m/M)^{1-\alpha}$ and the dimensionless time $\tau = -\ln\left[1 - (m_0/M)^{1-\alpha} e^{-\gamma \alpha t}\right]$ fits the observations equally well for both values of $\alpha$ (see figure). For simplicity, we have chosen $M$ from Table 1 of ref. 1. We have determined $\gamma_\alpha = a_\alpha(1-\alpha)/M^{1-\alpha}$ using the values of $a_{3/4}$ from the same Table and $a_{2/3}$ for the cow, hen and guppy to be 0.62, 0.67 and 0.064 respectively in order to ensure that $\gamma_{2/3} = \gamma_{3/4}$ for simplicity. (We have not tried to adjust the value of $M$ and $\gamma$ to get a better fit because that is not the point of this comment.) Also, an equally good fit of the universal curve is obtained (see last panel of the figure) for the three species with both values of $\alpha$. Furthermore, the form of the universal curve (Eqn. (4)) is <u>independent</u> of the value of $\alpha$. Thus the existence of a universal curve implies nothing about the value of $\alpha$.

Jayanth R. Banavar\*, John Damuth†, Amos Maritan‡, Andrea Rinaldo§

*\* Department of Physics, 104 Davey Laboratory, The Pennsylvania State University, University Park, Pennsylvania 16802*

*† Department of Ecology, Evolution and Marine Biology, University of California, Santa Barbara CA 93106*

*‡ International School for Advanced Studies (S.I.S.S.A.), Via Beirut 2-4, 34014 Trieste, INFM and the Abdus Salam International Center for Theoretical Physics, Trieste, Italy*

*§ Dipartimento di Ingegneria Idraulica, Marittima e Geotecnica, Universita' di Padova, Padova, Italy*

Figure caption. Panels (A–C) show plots of growth curves. + = empirical data; X = the best fit obtained in Ref. (1) (i.e., $\alpha = 3/4$); * = plot of Eqn. (3) with $\alpha = 2/3$ and the values of $M$ and $\gamma$ as obtained in Ref. (1). Panel D shows the universal growth curve (Eqn. (4)) with data from the three species with both $\alpha = 3/4$ and $\alpha = 2/3$.

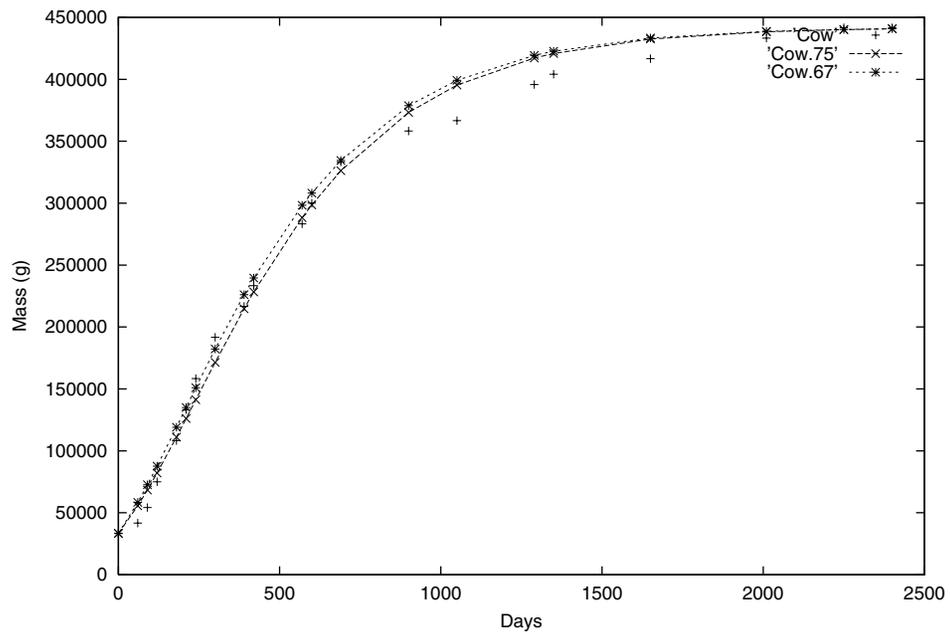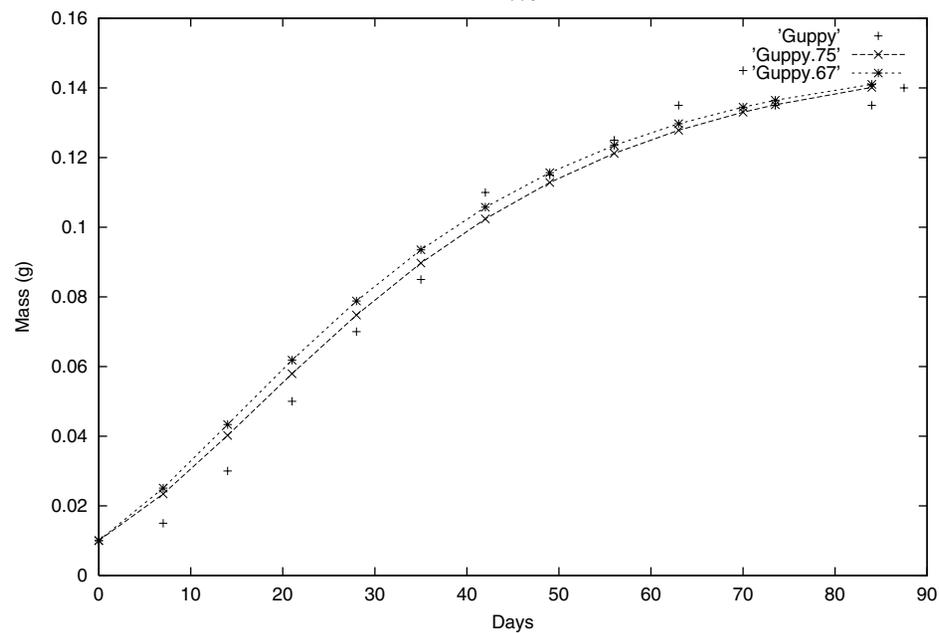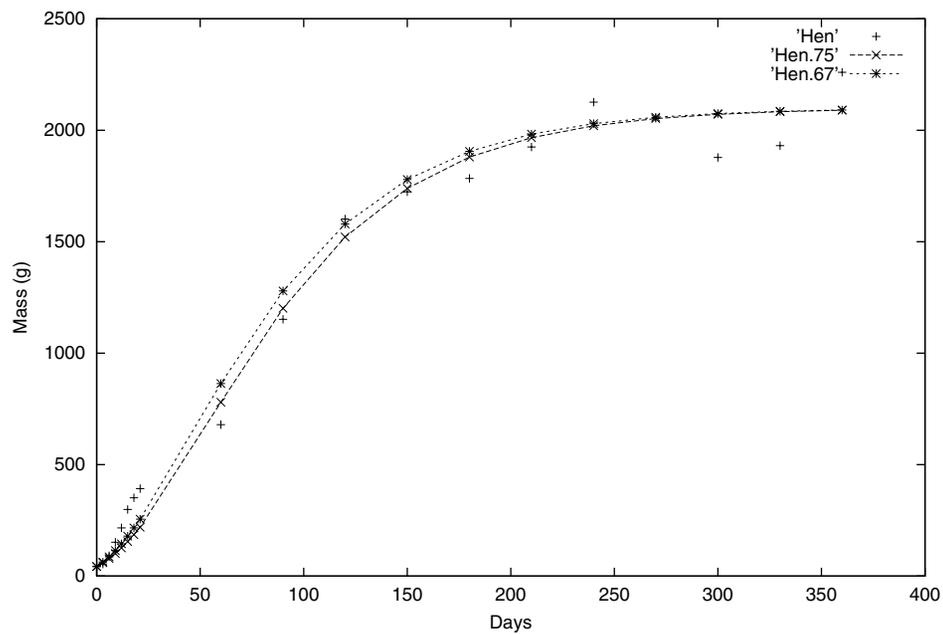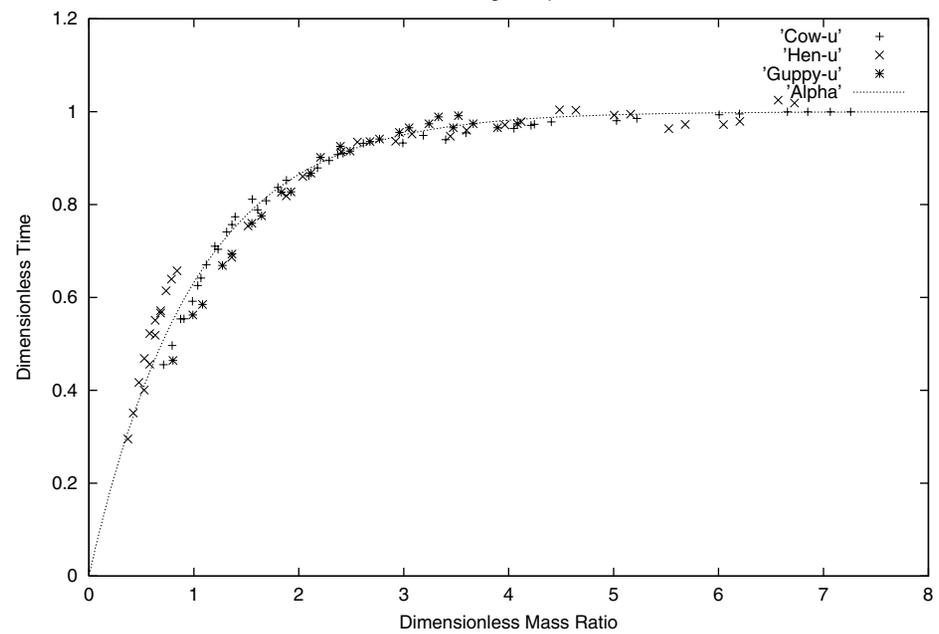